 \documentstyle[aas2pp4]{article}

\slugcomment{Not to appear in Nonlearned J., 45.}

\lefthead{Longbottom et al.}
\righthead{Flux Braiding}

\begin{document}

\title{Magnetic Flux Braiding: Force-Free Equilibria and
Current Sheets}

\author{A. W. Longbottom and G. J. Rickard\altaffilmark{1}}
\affil{Dept. of Mathematical and Computational Sciences,\\
University of St. Andrews, KY16 9SS, Scotland.}

\and

\author{I. J. D. Craig and A. D. Sneyd}
\affil{Dept. of Mathematics,
Waikato University, Hamilton, New Zealand.}


\altaffiltext{1}{present address: The Met Office, London Road, Bracknell, Berks, RG12 2SZ, UK}


\begin{abstract}
We use a numerical nonlinear multigrid magnetic relaxation technique
to investigate
the generation of current sheets in three-dimensional magnetic
flux braiding experiments. We are able to catalogue the relaxed
nonlinear force-free equilibria resulting from the application
of deformations to an initially undisturbed region of plasma containing
a uniform, vertical magnetic field. The deformations are manifested by
imposing motions on the bounding planes to which the magnetic
field is anchored. Once imposed the new distribution of magnetic
footpoints are then taken to be fixed, so that the rest of
the plasma must then relax to a new equilibrium configuration.
For the class of footpoint motions we have examined, we find that
singular and nonsingular equilibria can be generated. By singular
we mean that within the limits imposed by numerical resolution we
find that there is no convergence to a well-defined equilibrium
as the number of grid points in the numerical domain is increased.
These singular equilibria contain current ``sheets" of
ever-increasing current intensity and decreasing width; they occur
when the footpoint
motions exceed a certain threshold, and must include both twist
and shear to be effective. On the basis of these results we contend
that flux braiding will indeed result in significant
current generation. We discuss the implications of our results
for coronal heating.
\end{abstract}




%

\section{Introduction}
Accounting for the elevated temperature of the corona is a
fundamental issue in solar observation and theory. For many
it is the magnetic field that is at the heart of the mystery,
acting as the link between the convective motions in the
photosphere and the resultant currents and resistive dissipation
in the corona itself. Such a link was perhaps first developed
by Gold (1964), and subsequently formalised to some degree
by Parker (1972,1983) in his notion of ``topological dissipation"
to describe the process by which twisting and braiding of the ambient
magnetic field can lead to energy release in the corona. Indeed,
in his 1972 paper, Parker reduced the problem to that of considering
nothing more than a uniform field contained between a pair of
superconducting plates. This latter configuration has become known
as ``Parker's model", and has proved useful as a testing ground
for coronal heating theories.

The currents $j$ produced in the corona must be of significant
intensity to account for coronal heating. This is because the
coronal resistivity $\eta$ (typically of the order of
$10^{-12}$Mm$^{2}$s$^{-1}$) in the ohmic dissipation term
${\eta}j^{2}$ is so small that only currents of a sufficient
magnitude can lead to energy deposition within the required
coronal timescales. To produce such currents, the magnetic field
must be able to collapse to sufficiently small scalelengths.
Indeed, if we take the equilibrium field to contain scalelengths
of order unity, then we require a collapse by at least six orders
of magnitude to generate the intense currents. Such fine scale
current structures we shall refer to as ``current concentrations".
These are to be contrasted with ``current sheets" which are taken
to be those structures resulting from spatial discontinuities in the
magnetic field direction, and whose existence within the framework of
ideal magnetohydrodynamics (MHD) is generally considered to be
singular, in that it is only the presence of finite resistivity that
allows their spatial extent to be resolved. Current concentrations,
on the other hand, can be resolved within ideal MHD. However we
choose to define these features, it is this vast range of scalelengths
they imply that certainly makes the task of finding evidence
for them so challenging.

From an analytical point of view, Parker's model (1972) contends that
current sheets will in general be created in coronal magnetic fields.
This is based on Parker's belief that equilibrium is only possible if
an ignorable coordinate exists in the magnetic field perturbations.
Given the random nature of photospheric flows, it seems unlikely
that the perturbations will possess a certain symmetry. Therefore, in
order to proceed towards an equilibrium, the magnetic field must regain
this symmetry, and to do so the field must alter its topology. Parker
suggests that this is accomplished through current sheets and magnetic
reconnection. Since the topology is changed through the nonidealness of
the plasma, the term ``topological dissipation" was coined.
However, van Ballegooijen (1985, 1988a) has pointed out a flaw in
Parker's original analysis, and claims in his own right that the field
can pass through a series of equilibria without requiring reconnection.
These equilibria are continuous and have no symmetry.
Hence it is the time evolution through these equilibria that can lead
to coronal heating via current concentrations, and not as a result of
the coronal field attempting to relax to any particular equilibrium state.
The jury is still out on this issue, however, as evidenced by Parker's
recent work (1994).

Numerical attacks have been made on the problem, notably
van Ballegooijen (1988b, 1990), Miki\'{c} et al (1989), Longcope and
Sudan (1994), Galsgaard and Nordlund (1996) and Hendrix and Van Hoven
(1996). On the evidence from
such modelling it does seem plausible that magnetic scalelengths much
smaller than those proscribed for the imposed flows can result in the
corona, thereby lending weight to this process being a coronal heating
mechanism. Furthermore, the work of Miki\'{c} et al (1989) shows that
force-free equilibria without current sheets can be generated for each
particular distribution of footpoints comprising the overall evolution.
Longcope and Strauss (1994a,b) have examined the formation of current layers
with a nonzero, but small, thickness. They show, through the analysis of
of the Jacobian of the fieldline mapping, that these current consentrations
may easily be more than six orders of magnitude smaller than the equilibrium
length scale. The resulting structures they observe bear a marked resembelance to those we
find in this paper, albeit through the use of different methods to those presented here.
On the basis of these few experiments it would seem that the weight
of evidence is against Parker. However, as we have previously noted, the
coronal scalelengths actually envisaged are so small as to be out
of the reach of such numerical simulations for the foreseeable future.
We must therefore tread carefully in extrapolating the numerical
results to actual coronal conditions.

In order to try to alleviate the numerical shortfall, resulting
from the severe scalelength constraints, Craig and Sneyd (1986) (1990)
developed a magnetic relaxation technique using a Lagrangian
scheme. Such a scheme permits the numerical mesh to move with the
fluid elements and magnetic field lines comprising the plasma.
Consequently, as any short scalelengths develop in a region of the
plasma, so the Lagrangian scheme will aggregate more grid points there.
This means that for the same number of initial numerical nodes as an
Eulerian code might have, the Lagrangian scheme will be able to
resolve much finer scales. This is clearly an advantage when the search
is directed specifically towards finding evidence of current
concentrations or sheets. We will therefore be using the Lagrangian
scheme in this study. This is not to say that the Lagrangian code is
a panacea for all our resolution ills, since we still need to have
sufficient resolution over the whole magnetic structure containing
any of these current structures. Even allowing for this latter
constraint, there is no doubting the efficacy of the
Lagrangian scheme in allowing us to explore a greater range of parameter
space than would be accessible to any Eulerian code for the equivalent
amount of computing power.

We wish, therefore, to directly test Parker's model of current sheet
formation using the Lagrangian relaxation scheme. The method allows
us to find the nonlinear, force-free equilibria consistent with the
imposed magnetic footpoint displacements. We are consequently only
interested in the final magnetic configuration within the constraints
of ideal MHD. This contrasts with most of the previous numerical
studies of this problem which were Eulerian and time-dependent.
It should be noted, however, that although in the absence of definitive
analytic and numerical diagnostics, we cannot be certain that a solution
with a singular current sheet is necessarily the end result of a given
numerical experiment, our results at least indicate those configurations
with the potential to develop field singularities.

The layout of the paper is as follows. In Section 2 we outline the basic
equations and method of solution, and detail the numerical experiments
that we have undertaken. Section 3 presents the results and analysis from
these experiments, and we finish with a discussion and some conclusions in
Section 4.

%
%
\section{Model and Numerical Details}
\subsection{Basic Equations}
For this particular study we adopt the low-$\beta$ approximation
for the solar corona and look for equilibrium solutions of the
equation,
\begin{eqnarray}
(\nabla \times {\bf B}) \times {\bf B} = 0,
\label{eq:fff}
\end{eqnarray}
where ${\bf B}$ is the magnetic field. These equations are nonlinear.
To find solutions of these equations we follow the relaxation method
detailed by Craig and Sneyd (1986) (1990) and use a momentum equation
which ignores inertial effects in favour of a dominant frictional
term proportional to the fluid velocity ${\bf v}$, and takes gas pressure
to be negligible in line with the low-$\beta$ approximation. This simply
leaves us with,
\begin{eqnarray}
{\bf v} = {\bf j} \times {\bf B},
\label{eq:momen}
\end{eqnarray}
where the current ${\bf j} = \nabla \times {\bf B}$. In this form the
momentum equation guarantees relaxation toward a state of lower magnetic
energy. The equation set is completed by noting that ${\bf B} / {\rho}$
satisfies,
\begin{eqnarray}
{D \over Dt}{ {\bf B} \over \rho} =
({ {\bf B} \over \rho} \cdot \nabla){\bf v},
\label{eq:induct}
\end{eqnarray}
where $\rho$ is the fluid density, $D / Dt$ is the so-called
convective derivative, and
\begin{eqnarray}
\nabla \cdot {\bf B} = 0.
\label{eq:diverge}
\end{eqnarray}

The details for the solution of these equations using Lagrangian
coordinates can be found in Craig and Sneyd (1986) (1990). The vital
step is the realisation that the whole problem can be formulated in
terms of the Lagrangian position of a fluid element ${\bf x}$, say,
by replacing ${\bf v}$ by $D{\bf x} / Dt$, and using the so-called
Cauchy solution for equation (\ref{eq:induct}) which expresses
${\bf B} / {\rho}$ in terms of the equilibrium magnetic field ${\bf B}_0$,
density ${\rho}_0$, and ${\bf x}$ itself. The resultant system of
equations is parabolic, as shown by Craig and Sneyd (1990). As a
consequence, unconditionally stable numerical schemes must be used to
solve the system if inefficiently small time steps are to be avoided.
The implicit formulation devised by Craig and Sneyd (1986) gives the
requisite numerical stability, and can be solved using alternating
direction implicit (ADI) techniques.
It should be noted that although Lagrangian,
the equations themselves are solved on a fixed grid using centered
differences.

As for all classical methods, the number of iterations required for
convergence scales as some power of the number of points ($N$) in the
system, typically as $N^2$. Thus for numerical schemes in three
dimensions, doubling
the size of the computational domain requires approximately 64
times the number of iterations. It was found that for this simulation
grids of over $32^3$ required an excessively long computational time.
In order to somewhat alleviate this problem we have implemented a
nonlinear multigrid algorithm for grid sizes of $32^3$ or more.
For such a scheme the number of iterations scale at a fraction
of that of classical methods thus allowing high resolution
calculations to be conducted in realistic times. Our method is
outlined in the appendix.

\subsection{The Numerical Experiments}

Our equilibrium configuration is exactly that envisaged by Parker
(1972), i.e., a uniform, vertical magnetic field
${\bf B}_{0}=B_{0}{\hat{\bf z}}$, with its ends anchored into superconducting
plates at $z = \pm L_z$. The initial positions of the fluid elements are
distributed uniformly throughout a domain bounded by the superconducting
plates in $z$, and by distances $x = \pm L_x$ and $y = \pm L_y$. In the
previous experiment of the nonlinear evolution of the kink instability
detailed by Craig and Sneyd (1990), the fluid elements lying in the
bounding planes in $x$ and $y$ were taken to be fixed. Therefore only
fluid elements inside the computational domain were free to move. In our
initial experiments we adopted the same boundary conditions. However, the
nature of our experiments meant that we generated boundary layers at the
$x$ and $y$ bounding planes which tended to mask the physics of importance.
To overcome this problem, the $x$ and $y$ directions are taken to be
periodic in the displacements of the fluid elements, leaving the only
fluid elements fixed to be those in the planes $z = \pm L_z$. Apart from
the initial equilibrium, this is the only major change to the classical
Lagrangian code from its previous incarnation under Craig and Sneyd (1990).
However for the systems with large shears and large numbers of grid points
a nonlinear multigrid scheme has also been implemented (see Longbottom,
Fielder and Rickard (1997) for further details).

Starting from the equilibrium distribution just detailed, we impose
displacements of the fluid elements within the planes $z = \pm L_z$.
These displacements take the form of shears of equal magnitude but opposite
direction on the two driving boundaries.
Once displaced, these elements are then held fixed, and the remaining
fluid elements are allowed to relax towards a new equilibrium
configuration consistent with these new boundary conditions. We say that
we have a converged solution when both our norm of the residual force
has fallen to at least $O(10^{-3})$ (starting at $O(10)$) and the maximum
current has converged to two decimal places.
By solving the ideal MHD
equations, we guarantee that the magnetic field lines are ``frozen-in''
to the fluid. The equilibrium field lines in this case are vertical
lines running from the bottom to the top boundary in $z$, so that fluid
elements initially sharing the same values of $x$ and $y$ lie on the same
field line. As we have intimated, they will lie on the same field
line throughout the relaxation. We can therefore plot any relaxed field
line simply by drawing a line through that same set of fluid elements.
This results from our choice of equilibrium, and using the Lagrangian
coordinate system.

For every single solution, we attempt to detail numerical convergence in
the solution by repeating the same relaxation with ever increasing numbers
of fluid elements. In this way we hope to separate out equilibria that
are non-singular (smooth) from those that are singular, i.e., containing
current sheets. If the equilibrium is singular, then increasing the
numbers of fluid elements will not result in a convergent maximum current
in the current sheet. The scalelength of such a singular feature will always
be smaller than that attainable numerically, and so the current locally will
continue to increase. However, the total current in the computational
domain will converge, showing that the global solution is convergent, apart
from the locality of the singularity. On the other hand, the nonsingular
solutions will converge to a smoothly resolved structure throughout the
whole domain. Even with the Lagrangian code, the ultimate lack of
numerical resolution means that the convergence properties of any equilibrium
solution are all that we can base our physical results upon. The
relaxation experiments described by Billinghurst et al (1993) to examine
current sheet formation in two-dimensional fields provide a substantive
example of this process, and it is indeed their convergence tests that
we have employed here. It should be noted that although increasing the
number of grid points reduces the numerical error for a particular
solution, the solution itself depends on the position of the Lagrangian grid
points. Within the confines of one dimensional theory it can be shown
that the motion of the Lagrangian grid allows a better representaion of
the true solution as compared to an Eulerian grid with the equivalent
number of grid points. The maximum current that may be resolved scales
as the square of the number of grid points. The behaviour in a higher
number of dimensions is not so obvious, but for many cases an improved
representation of the true solution over the equivalent Eulerian code
is found.

Having outlined the basis for our numerical experiments, we now need to
consider possible footpoint motions. The key ingredient is to produce
scalelength collapse of the magnetic field within the volume as a
result of smooth displacements on the boundary. This is the crux of
Parker's model (1972). Clearly, fine scale motion on the boundary will
manifest itself as fine scales within the volume. However, the observations
suggest that the scale of convective motion in the photosphere is much
greater than we envisage for the current sheets. It must therefore be
left up to the coronal magnetic field to produce the short scalelengths
of its own accord. Parker (1972) believed that the coronal magnetic field
always does, and hence the notion of topological dissipation.

Our particular choice of footpoint motions are motivated by the work
of Galsgaard and Nordlund (1996). They conclude that the most important
factor leading to current sheet formation is one shear in one direction
followed by another in the orthogonal direction. This process is
deemed to be sufficient to produce the exponential current growth
predicted by van Ballegooijen (1986) on the basis of a random series
of shearing motions. As Galsgaard and Nordlund (1996) note, it is the
underlying field and flow topology resulting from the shear plus a
shear that inevitably leads to the exponential growth. It seems clear
that the alternating shear flow profiles advocated by
van Ballegooijen (1988a), and tested by van Ballegooijen (1988b) and
Miki\'{c} et al (1989), will still lead to the exponential growth, since
the sequence of footpoint motions is indeed composed entirely of the basic
element, i.e., a shear plus a shear. To simplify the problem as much as
possible, then, we choose to examine the possible effect on the coronal
field of a shear in one direction, followed by a shear in the same, or
in another, direction. It is these basic elements that we are going to
focus on here.

In the numerical experiments, distances are normalised to the initial
length of each field line, and we take the normalising magnetic field
strength and plasma density to be those in the equilibrium. For the
experiments we have conducted to date, we use the unit cube as the
initial volume of our numerical domain, and we take the field strength
to be one and density to be 0.05. These provide the reference points when
we later detail results from the relaxed equilibria.

To maximise the spatial resolution available to us, we will simply
examine the equilibria associated with the footpoint displacements shown
in Figures~\ref{footd}(a)-(f). Figure~\ref{footd}(a) shows the initial
equilibrium footpoints. At every grid point (at the intersection of
grid lines) lies the footpoint of each magnetic field line. The Lagrangian
code follows these particular points, so that after each successive
distortion of the grid we can exactly identify where the original
footpoint locations shown in Figure~\ref{footd}(a) have been moved to.
Figure~\ref{footd}(b) results from a
shear of magnitude $0.8$ applied parallel to the $y$-axis.
The remaining Figures~\ref{footd}(c)-(f) result from shears of magnitudes
$0.1$, $0.3$, $0.5$, and $0.7$, respectively, applied parallel to the
$x$-axis {\it after} having previously applied the shear resulting in
Figure~\ref{footd}(b). We are therefore examining the effect of shearing
in one direction, immediately followed by shearing in the perpendicular
direction. We hope to show that this is sufficient to produce scalelength
collapse within the plasma volume. It should be noted that the displacements
on the boundaries $L=\pm L_z$ are of the same magnitude but of opposite
sign.

From the evidence of some of the gross grid distortions shown
in the previous Figures, the reader may be surprised that we
can claim to obtain relaxed equilibria from such complex arrangements.
We should point out, however, that we have shown the physical grid
displacements, and not the uniform grid on which the resulting Lagrangian
equations are actually solved. As we shall show, the more distorted
the system, the more difficult the relaxed equilibrium is to obtain.
Nevertheless, given sufficiently robust relaxation techniques, the
Lagrangian system should allow us to approach the equilibrium
associated with each set of footpoints arbitrarily closely.
\placefigure{footd}

Although we have detailed the footpoints, there is still the question about
what to use as the starting points for the fluid elements within the
computational volume. As Craig and Sneyd (1990) have intimated, we have
a nonlinear equation to solve, and it is therefore possible that differing
initial conditions for the same footpoint distribution will result in
different relaxed states. While it would be interesting to explore this
hypothesis, we are faced with the more practical problem of obtaining
any relaxed equilibria at all. This is because not all initial conditions
lead to solutions, despite the unconditional numerical stability of the
Lagrangian code. It seems that sometimes the code finds an initial
condition too complex to be able to sensibly unscramble, and fails. The
best strategy we have found to date is to use as an initial guess the
relaxed solution from the previous state in the sequence of footpoint
displacements, with the new footpoint displacements prescribed on the
boundaries $z=\pm L_z$. So, although we view each equilibrium as an entirely
separate solution to all the others, there is a sense in which they can
be viewed as a sequence connected together. The ``dynamical'' experiment
of Miki\'{c} et al (1989) falls into this category.

Having applied the footpoint displacements to the end planes at
$z={\pm}L_z$, these are then held fixed, and the remaining fluid
elements are able to move in such a way as to approach a new
equilibrium consistent with the footpoint distribution. How close
we are to the new equilibrium is measured by the maximum Lorentz
force ${\bf j} \times {\bf B}$ within the domain. Having obtained
a ``suitably good'' solution, we then calculate the maximum
current within the domain, and the total integrated current strength
$|\bf{j}|$ over the
whole domain. We then repeat the whole experiment for increasing
numbers of fluid elements within the domain. The whole execise is
then repeated for each new distribution of footpoints. Based on these
results, we are then in a position to comment on the smoothness or
otherwise of a given equilibrium solution.

Our ability to catalogue each equilibrium accurately is limited
firstly by the available computing power, and secondly (and perhaps
more importantly) by the speed at which the equilibrium solution is
approached numerically. The last feature is compounded by increasing
numbers of fluid elements and the increasing complexity of each
equilibrium solution. This means that, even with the implementation
of multigrid techniques, the equilibria
associated with the most distorted grids are not only the
hardest to solve computationally, but also, because of their inherent
problems, are the ones on which we have the least amount of
numerical convergence.

%
\section{Results}

The summary of all our results is encapsulated in Figure~\ref{jmod}.
This plots the maximum absolute value of the current $J_m$ obtained
within the computational domain (always at the centre of the mid-plane
in $z$) in the best relaxed solution, against a measure of
the number of grid points in the simulation $n_a$. Here the total number
of points in the simulation is equal to $n_a^3$, and the measure
of the best relaxed solution is that with the lowest maximum value of
the Lorentz force ${\bf j} \times {\bf B}$ obtained so far. The number
labelling each curve shows the amplitude of the shear applied in the
$x$-direction, while the unlabelled curve refers to that in
Figure~\ref{footd}(b). Converged solutions (either using the classical
relaxation method or nonlinear multigrid) are plotted with an $\ast$,
those that are under estimates of the current are plotted with a $\triangle$.
\placefigure{jmod}

Figure~\ref{jmod} reveals that $j_m$ is tending to saturate for the lower
set of shears applied, i.e., up to and including $0.5$. We would therefore
conclude that these reveal smooth equilibria. However, for the shears
labelled by $0.6$ and $0.7$ the value of $J_m$ is tending to increase
at least linearly with $n_a$, so that within the confines of our
resolution we would conclude that these point to the possibility of
(at least) significant current concentrations, if not current sheets
themselves. It should be noted that, within the confines of a one
dimensional Lagrangian representation, a rigorous condition for a truly
singular current sheet may be obtained. This states that the maximum
current $J_m$ must scale as $n_a^{2}$. If this is taken as a generally
necessary condition for current sheet formation, then only equilibria with
a shear of $0.7$ exhibit singular behaviour, $J_m$ scaling approximately
linearly with $n_a$ for a shear of $0.6$. It is not however obvious
how the effect of working in a higher number of dimensions would
modify the condition. Thus we refer to both shears of $0.6$ and $0.7$ as
divergent solutions.
\placefigure{idlx}

Although we have a broad range of footpoint distributions, the generic
forms of the relaxed solutions we see divide themselves into two types,
at least with respect to the structure of the current. The first is
associated with shearing only in the $y$-direction. Here there is no
specific peaking of the current distribution. Rather we have a relatively
broad current structure filling most of the domain. As Figure~\ref{jmod}
shows, we require only relatively modest numbers of fluid elements
before we confidently predict that the equilibrium is a convergent
one with respect to $n_a$. Indeed, for certain parameters, this equilibrium
may be calculated analytically (Hood, priv. comm.). For these cases the
numerical method reproduces the true solution to within the given tolerance.
The basic field structure as a function
of $x$ of these solutions is shown in Figure~\ref{idlx}.
Figure~\ref{idlx}(a) shows the variation of $B_x$ with the height
$z$. The bold curve with the largest amplitude is $B_x$ close to
the lower $z$ boundary. As you go up in $z$ the amplitude of $B_x$
gradually falls, resulting in the second bold curve plotted at about
half way between the lower $z$ boundary and the mid-plane. In the
mid-plane, $B_x$ is zero. From the mid-plane upwards the variation
in $B_x$ is a mirror image of that below the mid-plane, with the
sign of $B_x$ reversed (as indicated by the dashed curves). There is
only a single curve shown for $B_y$ in Figure~\ref{idlx}(b) as
the relaxed $B_y$ is practically uniform in $z$. The variation in
$B_z$ is similar to that for $B_x$, i.e., a weak boundary layer at the
top and the bottom, with hardly any $z$-variation about the mid-plane.
The only difference is that there is no sign change in $B_z$, hence the
two curves, the larger amplitude one in the mid-plane, and the other
close to the $z$-boundaries.

We also note that the variation in $x$ of $B_y$ exactly matches of
that of the imposed footpoint displacement, i.e., proportional
to ${\rm sin}(2{\pi}x)$, whereas $B_z$ has half the wavelength. This
variation in $B_z$ results from the need to balance the excess magnetic
pressure in the $y$-direction produced by the footpoint shear. $B_z$
therefore has to match the $x$-gradient of ${B_y}^2$, resulting in
the $x$-profile we see. To produce this $B_z$ profile the fluid elements
have to displace themselves in the $x$-direction in order to generate
regions of either increased or decreased $B_z$. Since our fluid elements
are tied to the field lines, this $x$-motion naturally results in a
$B_x$ that has the same wavelength as $B_z$, but is ${\pi}/2$ out of
phase.

These first type of solutions are therefore almost wholly dependent
on $x$ alone about the mid-plane, with most of the variation in $z$
occurring towards the top and bottom planes. The lowest energy state
for our boundary conditions seems to be one in which an almost
one-dimensional structure extends over most of the domain, with most
of the stress taken up in the weak boundary layers. This is reminiscent
of results reported by Browning and Hood (1989) and references
therein in the context of twisted flux tubes in which, as here, the
equilibrium is practically one-dimensional everywhere, except for
boundary layers near the line-tied boundaries. This seems to be the
lowest energy configuration with respect to one in which, as might have
been anticipated, there is a gradual variation in the properties of
the equilibrium over the whole length of the structure.
\placefigure{slicerz}

The second (and most important) type of current structure is
revealed in all the other relaxed solutions, once a second shear in
the $x$ direction has been applied. The only thing that
distinguishes each of these particular solutions is the amplitude
of the current obtained. Each solution contains a strong current
concentration restricted to a narrow slice passing through the centre
of each $z$-plane, which then rotates slowly as increasingly higher
$z$-planes are passed through. The amplitude of the maximum current in
each plane in this feature is practically constant along its length,
peaking in the central plane at $z=0$, and diminishing only very
gradually towards the two ends. The main component of the current is
in the $z$-direction. A representation of this feature is shown in
Figure~\ref{slicerz}, which shows the isosurface at $50$ per cent of
the maximum current in $j_z$. Larger amplitudes of $j_z$ are enclosed
within this isosurface, and this surface therefore reveals the extent
of the current concentration (and consequently its halfwidth, that we
define later). The effect of the shear is apparent in the twist of
the structure. We also note that there is just a single isosurface at
this level in the volume.
\placefigure{slicerx}

Representative isosurfaces in the same relaxed solution for the
currents $j_x$ and $j_y$ are shown in Figure~\ref{slicerx}(a) and
Figure~\ref{slicerx}(b), respectively. Comparing with
Figure~\ref{slicerz} we see that the isosurfaces for $j_x$ and $j_y$
exhibit the same twist, but that instead of a single isosurface we now
see the presence of two isosurfaces that, to a first approximation,
tend to wholly enclose the $50$ per cent isosurface for $j_z$.
In common with $j_z$, however, $j_y$ shows a fairly uniform spatial
distribution in $z$, whereas $j_x$ has a broader distribution near
the bounding planes in $z$, which then narrows and rotates as the
mid-plane is approached from either side.
\placefigure{grahamr}

The explanation for the second type of current structure is fairly
simple, and has been detailed by Galsgaard and Nordlund (1996). It all
boils down to how the field lines wrap themselves around one another
as a result of the two sets of shears, and the fact that they are
line-tied at each end. A schematic of the result of wrapping the
field lines is given in Figure~\ref{grahamr}, which shows the two
planes to which the footpoint displacements have been applied, along
with the resultant distribution of field lines (labelled with arrows).
The field line about which all the others wrap is shown as the vertical
dashed line. The bold field lines lie wholly in front of the vertical
line, while the dotted field lines lie wholly behind both these and
the vertical field line. The result of the magnetic tension and the
line-tying forces the field lines to compress on top of one another,
producing the strong currents that we see. In particular we note that
despite the shear the main component remains that in the $z$-direction.
Hence the single isosurface encompassing the current layer shown in
Figure~\ref{slicerz}. However, the components $j_x$ and $j_y$ retain
information relating to the imposed shears, and hence have a
change of sign on passing through the current layer in $j_z$. Therefore,
$j_x$ and $j_y$ have local maxima on both sides of the $j_z$ current layer,
and this explains the fact that they each reveal isosurfaces either side
of this layer.
\placefigure{jx}

While the isosurfaces reveal the general structure of the equilibrium
current, we have lost a sense of direction in them. This can be
recovered by considering some specific contours. In particular,
Figure~\ref{jx} shows contours of $j_x$ in planes of constant $z$
starting from the bottom plane, and working up to the mid-plane. There
is no need to go further in $z$ since, as the isosurfaces show, the
structures simply continue to rotate from those below the mid-plane,
albeit in the opposite direction. The dotted contours are for negative
current values. We now clearly see the effect of shear, with $j_x$
reversing direction about a specific line below the mid-plane, and
actually reversing about the line $y=0$ in the mid-plane itself. For
$j_x$, then, the structure in $z$ is tantamount to taking the mid-plane
contours and slowly ``sliding'' them past one another as you descend
or ascend through the planes.
\placefigure{jy}

The equivalent contours for $j_y$ are shown in Figure~\ref{jy}. The
mid-plane contours now reverse across the line $x=0$. As you move
through the planes in $z$ the contours rotate together, unlike
the ``sliding'' observed in $j_x$. The sets of contours for $j_x$
and $j_y$ not only have a certain spatial symmetry, but also their
amplitudes in that in each contour the maximum (positive) contour is
always the same magnitude as the minimum (negative) contour. The absolute
maxima are (a) 16.4, (b) 9.9, and (c) 5.3 for $j_x$ in
Figure~\ref{jx}, and (a) 7.8, (b) 10.2, and (c) 12.0 for $j_y$ in
Figure~\ref{jy}. We see that $j_x$ is tending to dominate $j_y$
close to the lower boundary, while the opposite is true by the time
we reach the mid-plane.
\placefigure{jz}

As we would anticipate from the isosurface plot, the contours of
$j_z$ shown in Figures~\ref{jz}(a)-(c) are dominated by the vertical
(positive) component. We also recover the sense of a single feature
that rigidly rotates as we progress from the lower to the upper boundary.
Specifically, the maximum and minimum contour
values are (a) 36.5 and -9.9, (b) 37.8 and -4.6, and (c) 37.5 and -3.5,
respectively. These contours also reveal how remarkably uniform the
current concentration in $j_z$ is with regard to its length and breadth
in each $z$-plane. We can therefore simply use the dimensions of the
current concentration in the mid-plane as a reference point.
Furthermore, since we obtain equilibrium solutions, then the
current ${\bf j}$ must be everywhere parallel to the magnetic field
${\bf B}$. Hence knowing the field line topology immediately gives us
the current topology.

As we have indicated, the main aim of these experiments is to determine
whether or not current concentrations or sheets are possible in these
sheared equilibria. Since the anticipated scalelengths are well below
those accessible numerically, it is only through tests of the numerical
convergence of our solutions that we can make predictions about the
behaviour of the actual coronal fields. It is therefore important to
have suitable reference points in each solution that can be reliably
measured, so as to allow the checks for convergence. Global checks of the
total current are one measure, as well as any local current maxima.
Here, the mid-plane solution provides such a useful reference point.
In particular, the current concentration associated with $j_z$ is
clearly produced by the shortening of the scales in the $x$ direction.
We will therefore use the distance across the line $x=0$ between the two
points where the current falls to half its peak value along the line $y=0$
as a relevant indicator of convergence.
\placetable{resolution}
\placefigure{cur_half}

Having defined this half-width length scale, we can map this on to the
the number of grid points that an equivalent Eulerian code would require
to gain the same resolution as the Lagrangian code. These are shown in
Table~\ref{resolution}. Obviously one could argue that an Eulerian code could
used a number of methods, for example a stretched grid, to gain more
resolution. However an equivalent method could normally be applied to the
Lagrangian code. For the largest shears we have been able, with the use
of the Lagrangian formulation coupled with nonlinear multigrid, to resolve
current structures of approximatley $1/1000th$ the size of the computational
domain. This is an order of magnitude better than an equivalent Eulerian code
(see for example Galsgaard and Nordlund (1996)). It should be noted,
however, that the Lagrangian formulation by definition can only examine
ideal MHD, and unlike resistive Eulerian codes nothing can be said about
reconnection or subsequent evolution once resistive effects come into play.
The resulting high resolution can however point to whether the current
structures seen in resistive codes are indeed ``current sheets'' or just
current concentrations that would allow slow reconnection at solar values
of resistivity. Here slow is with regard to observed timescales.

Figure~\ref{cur_half} shows a plot of maximum current against half-width
($\Delta$) as
the number of grid points is increased. Only the two shears that appear to
give a divergent current are shown. A tentative extrapolation may be
carried out to the width of a structure allowed by solar resistivity.
Fitting a power law to the data in figure~\ref{cur_half} gives the
relationship $j_m = 3.78/\Delta^{0.81}$. This corresponds, for a lengthscale
collapse of about six orders of magnitude, to a
current of the order of $3 \times 10^5 B_0/(\mu L_0)$ Am$^{-2}$, where $B_0$
and $L_0$ are characteristic magnetic field strengths and field line lengths.
Taking $B_0=10$G and $L_0=10^6$m this gives a current density of
$2 \times 10^6$Am$^{-2}$. If the energy within the current sheet is released
by Joule dissipation then this total energy may be calculated as
\begin{eqnarray}
j_D &=& \int j^2/\sigma \; dv
\end{eqnarray}
giving $j_D=4 \times 10^{18}$ J. For dissipation by reconnection acting on
the Alfven time scale (say the order of 100 seconds) we would get an energy
flux of approximately $4 \times 10^4$ Js$^{-1}$m$^{-2}$ over the area of the
box. Given the number of assumptions in making the above calculation, it is
still interesting to see that this value is comparable with energy flux
observed in active
regions of $10^4$Js$^{-1}$m$^{-2}$ (see for example Withbroe and Noyes (1977)).
\placefigure{conv}

A further measure of how close our solutions are to the true equilibrium
can be gained by plotting the integral of $|{\bf j}|$ over whole the box and
the magnetic energy within the box against the resolution used. Both of
these quantities should converge with increasing resolution for an equilibrium
solution, whether a current sheet is formed or not. We show these plots in
Figure~\ref{conv}. It can be seen, in all cases, that there is a smooth
convergence with increasing number of grid points.

%
%
\section{Discussion and Conclusions}

In this paper we have addressed the issue of the production of
current concentrations or sheets in nonlinear, force-free equilibria
in coronal magnetic fields. Previous work suggests that the combination
of a shear plus a shear is sufficient to do so, and we therefore
focus on such footpoint displacements. Our results to date show that
for certain amplitudes of the second shear, convergent, well-resolved
equilibria can be generated. However, beyond a critical point it seems
that the equilibria are divergent, in the sense that for increasing
numbers of fluid elements we find convergence in the total current and
magnetic energy in
the system, but that a local current peak (and its associated
scalelength) are not convergent. On extrapolating these latter results to
scalelengths expected in the coronal environment, we would predict
that currents of sufficient magnitude to account for significant
heating can be generated. Whether these equilibria are singular in the
sense of leading to current sheets we cannot determine. Nevertheless, the
current anticipated is sufficiently concentrated to make the point
academic; layers of resistively heated plasma will be present.

Furthermore, by use of a Lagrangian approach, we have been able to
access scalelengths well below those attained by any Eulerian code
to date. The Lagrangian code allows fluid elements to migrate into
those regions where scalelengths are reducing, and therefore makes it
particularly suitable for such equilibrium experiments. Field lines are
also simply plotted by tracing through the positions of the fluid elements;
there is no recourse to the use of integration routines etc. One
disadvantage of the classical relaxation method is its slow rate of
convergence for the most intensive solutions (in terms of complexity and number of grid points). Through the use of nonlinear multigrid methods we have been
able to partially overcome this problem. Thus converged solutions to these
most important experiments can be calculated in finite times.

If the reader is seeking a black and white answer to
whether Parker or van Ballegooijen is correct on the question of the
nature of coronal equilibria, then this is not the place to look.
There is certainly no degree of symmetry assumed in our simulations;
they are fully three-dimensional. Nevertheless our imposed footpoint
distributions have a certain symmetry to them. As for the equilibria,
we would contend that we find both well resolved, fully three-dimensional
solutions, as well as those that demonstrate a level of divergence.
This puts us firmly on the fence. Whatever one's views, we do present
compelling evidence for the generation of significant current
concentrations in the corona.

The question of the timescale required to produce such structures is
an important one. We only see the final product in terms of the
relaxed solution once the footpoint displacements have been imposed.
It must be left to others to determine whether a fully dynamical
corona can do likewise. Nevertheless, as pointed out by
Galsgaard and Nordlund (1996), the timescale for the footpoint
motions is a least a factor of a hundred less than that of the
characteristic Alfv\'en transit time, and one would therefore expect
a significant amount of time for near equilibrium conditions to evolve.
Future dynamical experiments with such separation of timescales will
have to be performed to demonstrate the truth or otherwise of this
assertion.

As ever, we have only begun to scratch the surface. Our bias towards
the Lagrangian approach ensures that we will continue to employ such
methods time dependently, hopefully to answer some of the questions of
timescale. We are also following up the work of van Ballegooijen (1988b)
with an Eulerian approach using the Euler potentials. With the advent of
nonlinear multigrid methods and more powerful workstations, it seems
entirely appropriate to see how far we can push the Eulerian codes in
this context!

\acknowledgements
The authors wish to express their gratitude to Klaus Galsgaard
for his prompting on the issue of flux-braiding and to the referee
for a number of useful comments.

\appendix
\section{The Implementation of Multigrid}

Here we outline our implementation of nonlinear multigrid with
respect to the above problem. More detailed accounts of multigrid
can be found in Brandt (1977), McCormick (1987) and for M.H.S.
relaxation theory Cally (1991), Fiedler (1992),
Longbottom, Fiedler and Rickard (1997).

The key to understanding multigrid methods is to know why classical
methods fail. Classical relaxation methods are extremely efficient
at smoothing out short wavelength errors, those on the scale of
the grid spacing. However they are poor at smoothing longer wavelength
errors. The trick with multigrid is to restrict the problem on to
successively coarser grids, smoothing the errors in each case, and then
prolongate the more accurate solution back on to the finer grids.
As smoothing on the coarser grids takes significantly less time than
smoothing on the finest grid such a multigrid sweep, smoothing all
wavelength errors, is much more efficient than a number of iterations
on the finest grid smoothing only short wavelength errors.

We implement a full approximation scheme (F.A.S.) fixed schedule
multigrid, including F-cycles. This proceeds as follows:

For solving $F({\bf x})={\bf j} \times {\bf B}=0$,
we start with a first guess for ${\bf x}$ (${\bf x}^{n}_F$) and (pre) smooth
on the fine grid to find the next best estimate ${\bf x}^{n+1}_F$.
We then compute the {\em residual} ${\bf r}_F=F({\bf x}^{n+1}_F)$,
and restrict both ${\bf x}^{n+1}_F$ and ${\bf r}_F$ to the next coarsest
grid by a second order three dimensional weighting
${\bf r}_C={\cal R}({\bf r}_F)$ and ${\bf x}^{n}_C={\cal R}({\bf x}^{n+1}_F)$.
Here $\cal R$ represents the restriction operator. The approximate solution to
the equation $F({\bf x}_C)={\bf r}_C + F({\bf x}^{n}_C)$,
(${\bf x}^{n+1}_C$) is then calculated on this coarser grid (by recursive
multigrid calls to successively coarser grids).
We then {\em prolongate} the correction to the solution,
${\bf x}^{n+1}_C-{\bf x}^{n}_C$, back up to the finer grid by trilinear
interpolation and add it to the best estimate on the fine grid
${\bf x}^{n+2}_F = {\bf x}^{n+1}_F + {\cal P}({\bf x}^{n+1}_C-{\bf x}^{n}_C)$.
Finally a further (post) smooth is carried out using ${\bf x}^{n+2}_F$
as the initial guess for the relaxation. This completes one multigrid sweep.

In general the efficience of the process is dependent on the order in which
the coarser grids are visited. We find, for this problem, that an F-cycle
is the best choice. Thus for a four leveled system (labeled 1 to 4 from
coarse to fine) one multigrid F-cycle visits the grids in the order
4-3-2-1-2-1-2-3-2-1-2-3-4.

Ideally for Multigrid the number of iterations should scale independently
from $N$ (the number of points on the finest grid), however in many complex
3D nonlinear cases the scaling is with some power of $N$ that lies
between 0 and 2. We find here that the number of iterations
scale like $N$ as opposed to $N^2$ for the classical method.
\placefigure{multigrid}

Figure~\ref{multigrid} compares the convergence of the classical and
multigrid versions of the ADI relaxation method. The residual force measures
the error in the solution over the whole domain. One work unit (WU is
the cpu time for one iteration on the finest grid. The examples shown
are for grids of $33^3$ and $65^3$ points after the second ($x$-axis)
shear of magnitude 0.2 has been applied. In both cases the convergence
of the classical method soon saturates (once short
wavelength errors have been smoothed), however the multigrid method
continues to converge. Also the scalings with number of grid points
for both classical and multigrid methods can be seen.

\clearpage
 
\begin{table*}
\begin{center}
\begin{tabular}{lccccccc}
     Lagrangian &\multicolumn{7}{c}{Effective Eularian resolution for
                                      given shear:}    \\
     resolution &  0.8 &  0.2 &  0.3 &  0.4  &  0.5  &  0.6  &  0.7   \\
\tableline
       $21^3$   &$26^3$&$39^3$&$42^3$&$ 46^3$&$ 48^3$&$ 50^3$&$ 52^3$ \\
       $31^3$   &$26^3$&$46^3$&$58^3$&$ 70^3$&$ 82^3$&$ 92^3$&$ 96^3$ \\
   $33^3\ast$   &$26^3$&$47^3$&$60^3$&$ 76^3$&$ 85^3$&$118^3$&$120^3$ \\
       $41^3$   &$26^3$&$50^3$&$68^3$&$ 92^3$&$134^3$&$200^3$&$308^3$ \\
       $51^3$   &$27^3$&$54^3$&$77^3$&$ 96^3$&$188^3$&$286^3$&$556^3$ \\
       $61^3$   &$27^3$&$58^3$&$85^3$&$133^3$&$234^3$&$462^3$&   --   \\
   $65^3\ast$   &$27^3$&$58^3$&$86^3$&$136^3$&$240^3$&$521^3$&$943^3$ \\
\end{tabular}
\end{center}

\tablenum{1}
\caption{The effective number of grid points needed by an Eulerian code
         to resolve the current structures obtained using the Lagrangian code.
         Resolutions marked $\ast$ were calculated using nonlinear multigrid.
 \label{resolution}}
\end{table*}

\clearpage

\figcaption[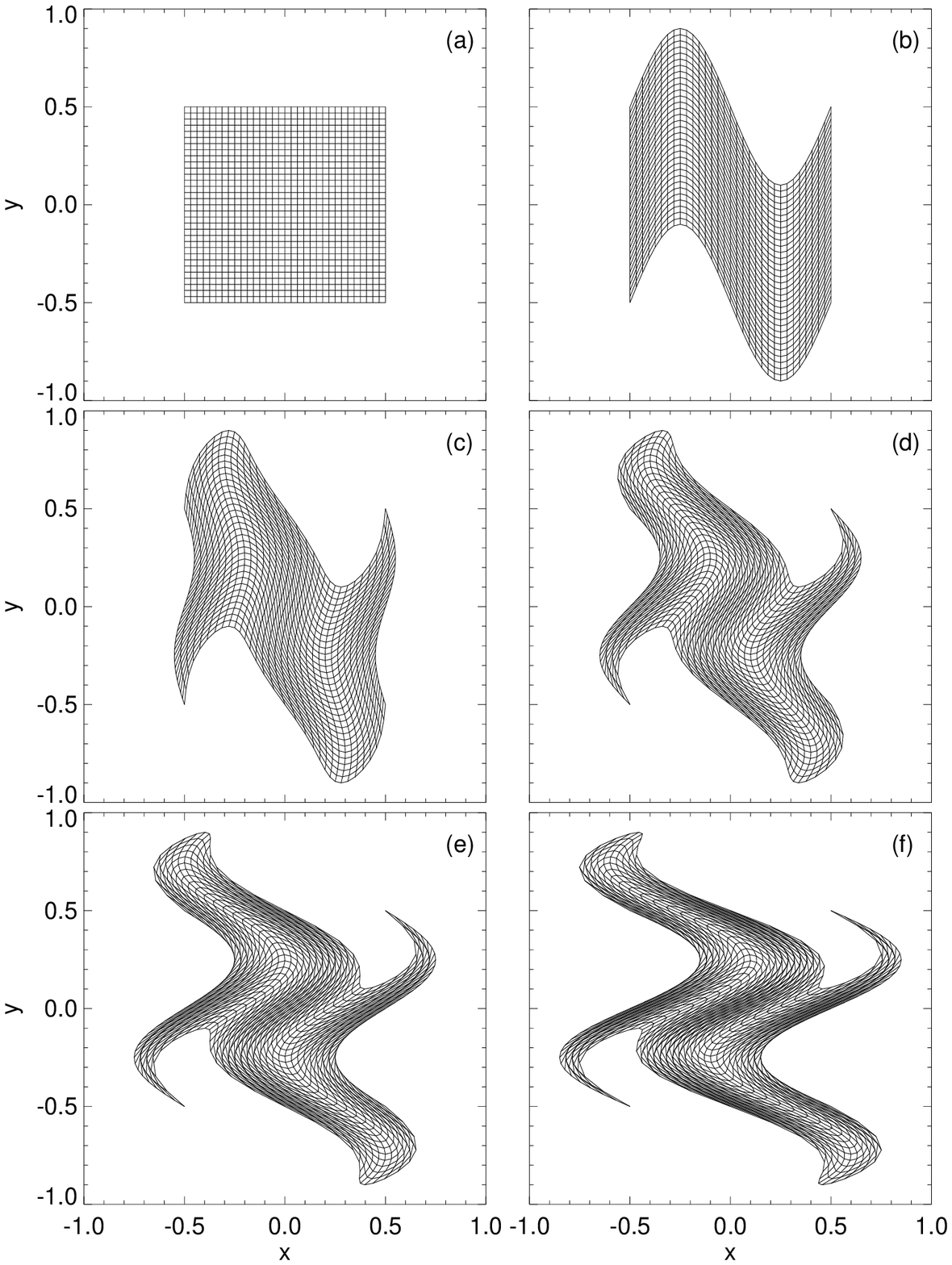]
{ Footpoint displacements applied on the plane
 $z=-L_z$. The displacements on the plane
 $z=+L_z$ are the mirror images of those shown,
 reflected in the planes $x=0$ and $y=0$.
 \label{footd} }

\figcaption[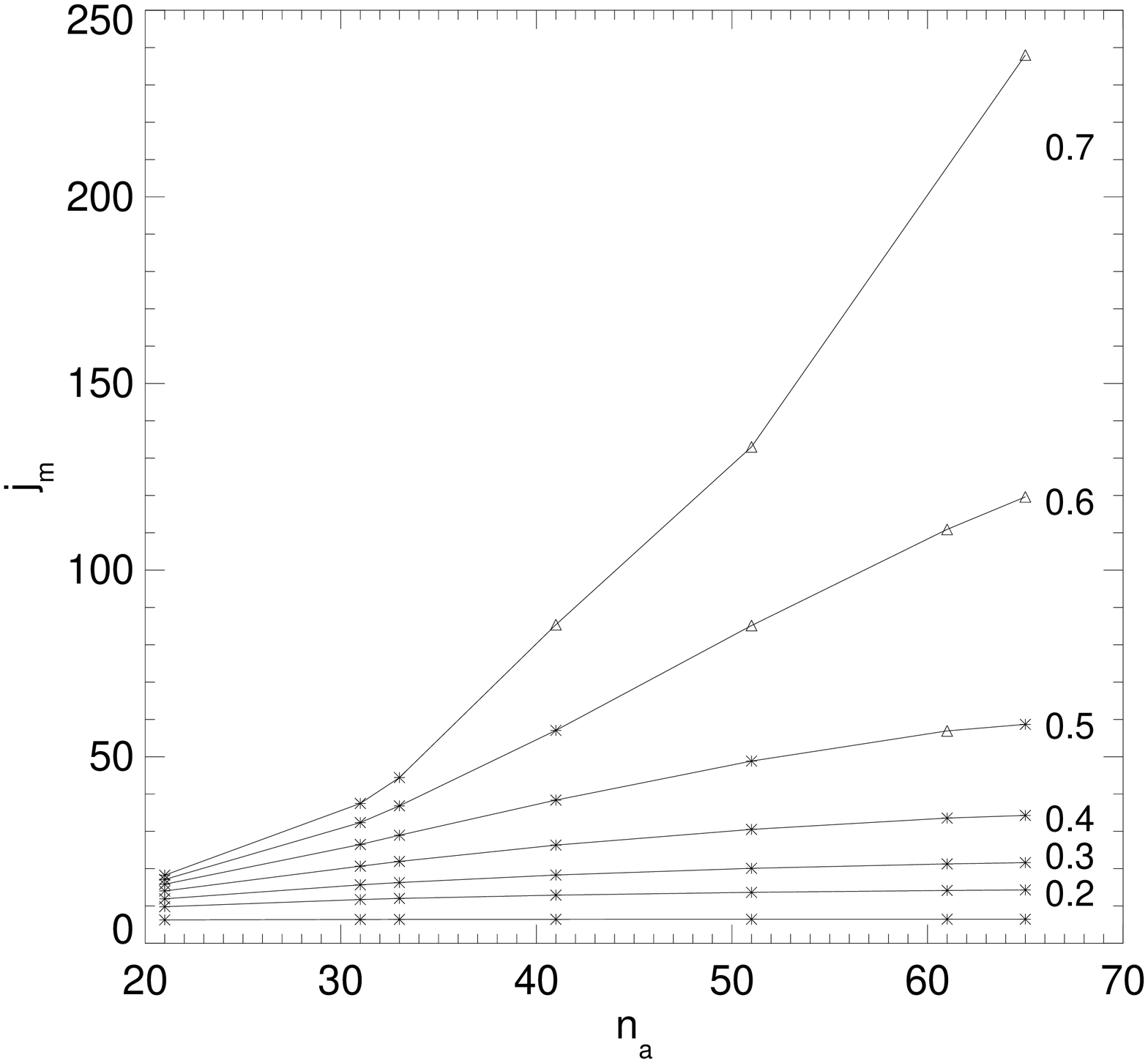]
{ Maximum current $j_m$ obtained in the best relaxed
 solution for each distribution of footpoints,
 plotted against $n_a$ (see text).
 \label{jmod} }

\figcaption[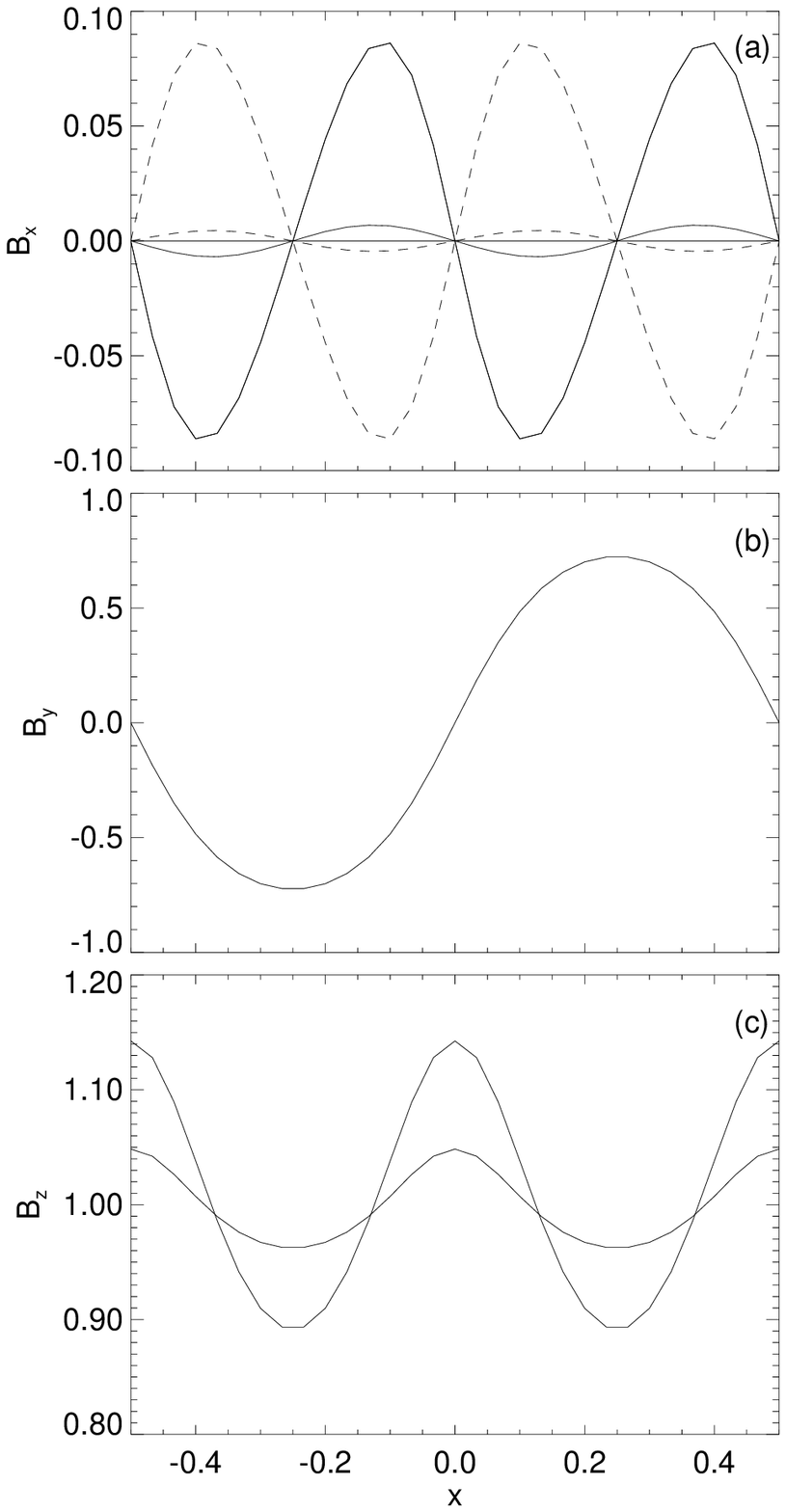]
{ Magnetic field structure for the first type of
 relaxed solution. The magnetic fields shown are
 (a) $B_x$, (b) $B_y$, and (c) $B_z$.
 \label{idlx} }

\figcaption[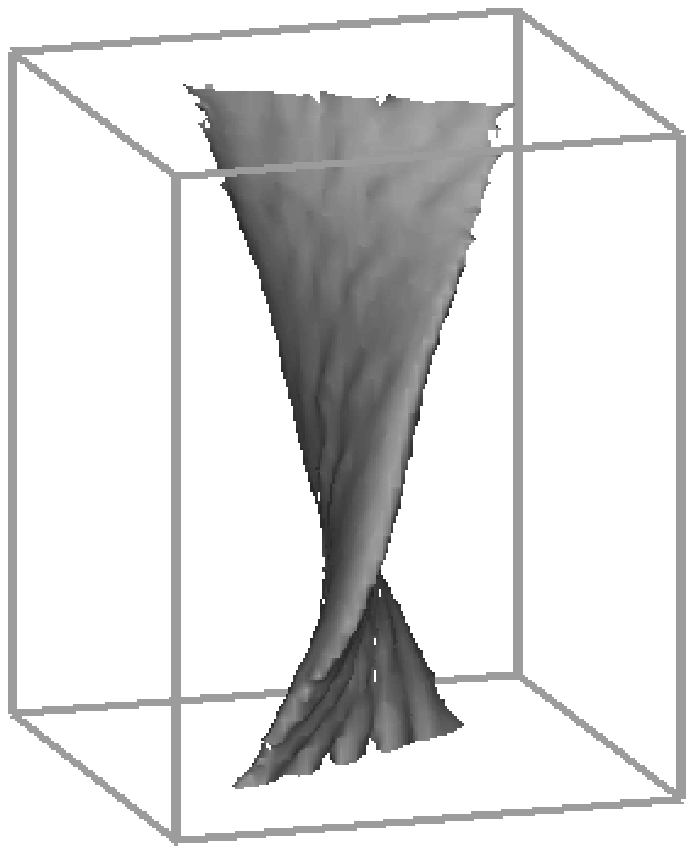]
{ Isosurface of the $j_z$ current component
 at $50$ per cent of the maximum current in
 the equilibrium solution. The vertical direction is
 $z$ going between $-0.5$ and $0.5$. The remaining
 directions are $x$ and $y$ forming a right-handed
 set and varying between $-0.3$ and $0.3$.
 \label{slicerz} }

\figcaption[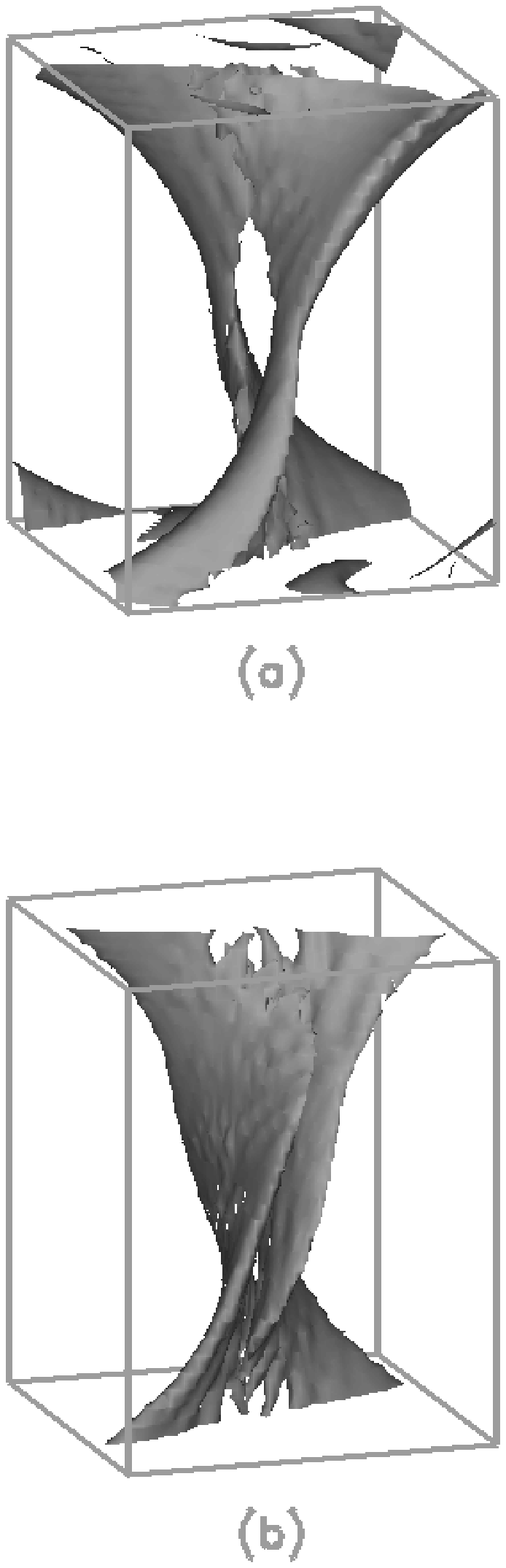]
{ Isosurfaces of  the (a) $j_x$, and (b) $j_y$
 currents at $25$ and $42$ per cent of the maximum
 currents for each component, respectively. The cube
 orientation and dimensions are as the previous Figure.
 \label{slicerx} }

\figcaption[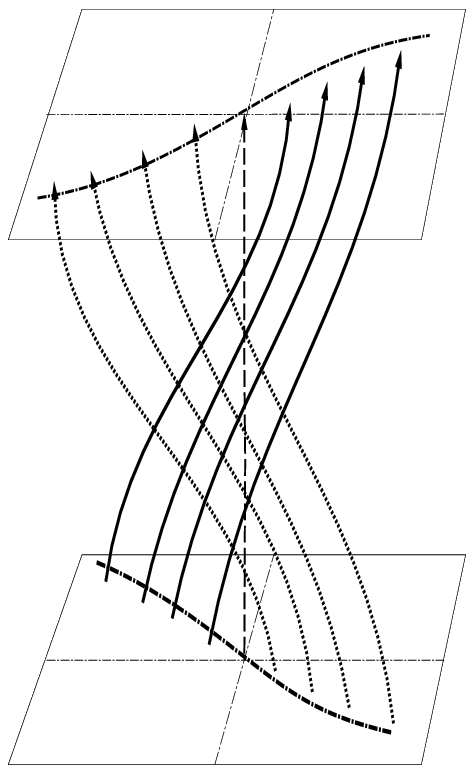]
{ Schematic of field-line wrapping resulting
 from the application of two shears. The field
 lines are labelled with an arrow.
 \label{grahamr} }

\figcaption[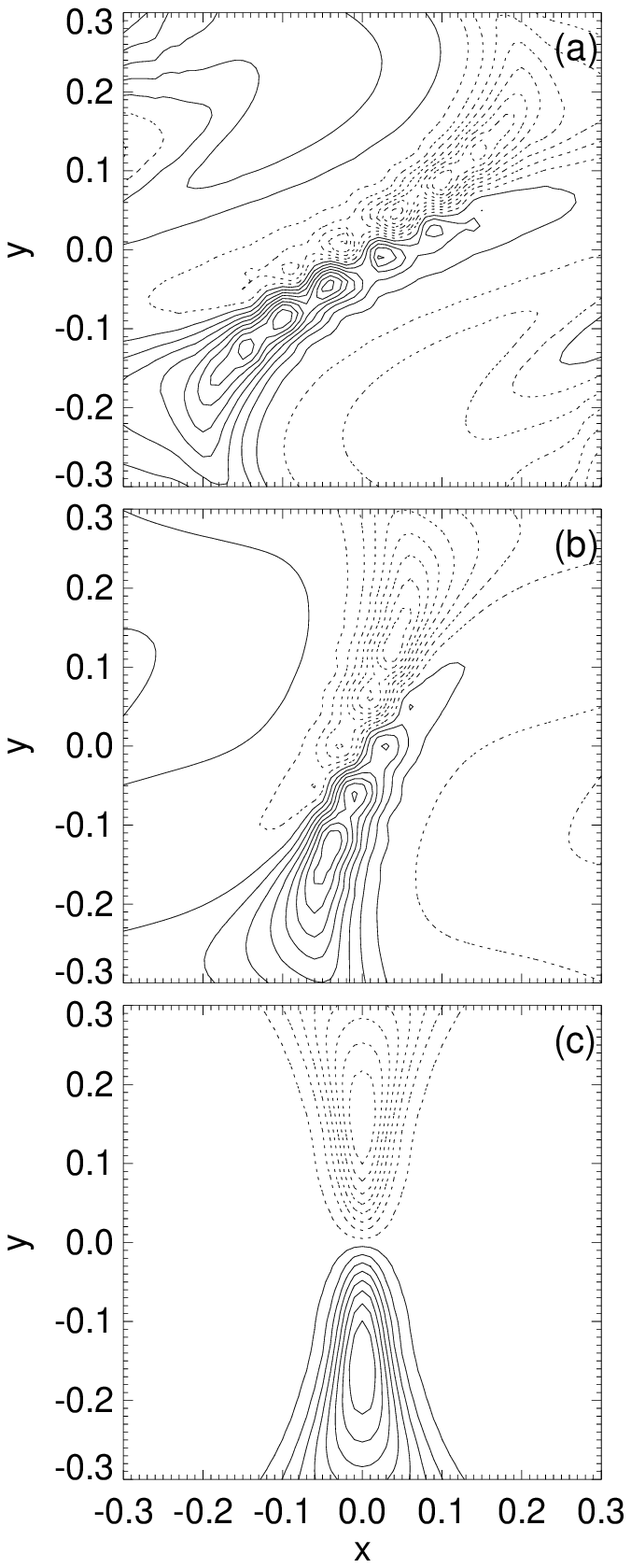]
{ Contours of the current $j_x$ in the planes
 (a) $z=-0.5$, (b) $z=-0.25$, and (c) $z=0$.
 \label{jx} }

\figcaption[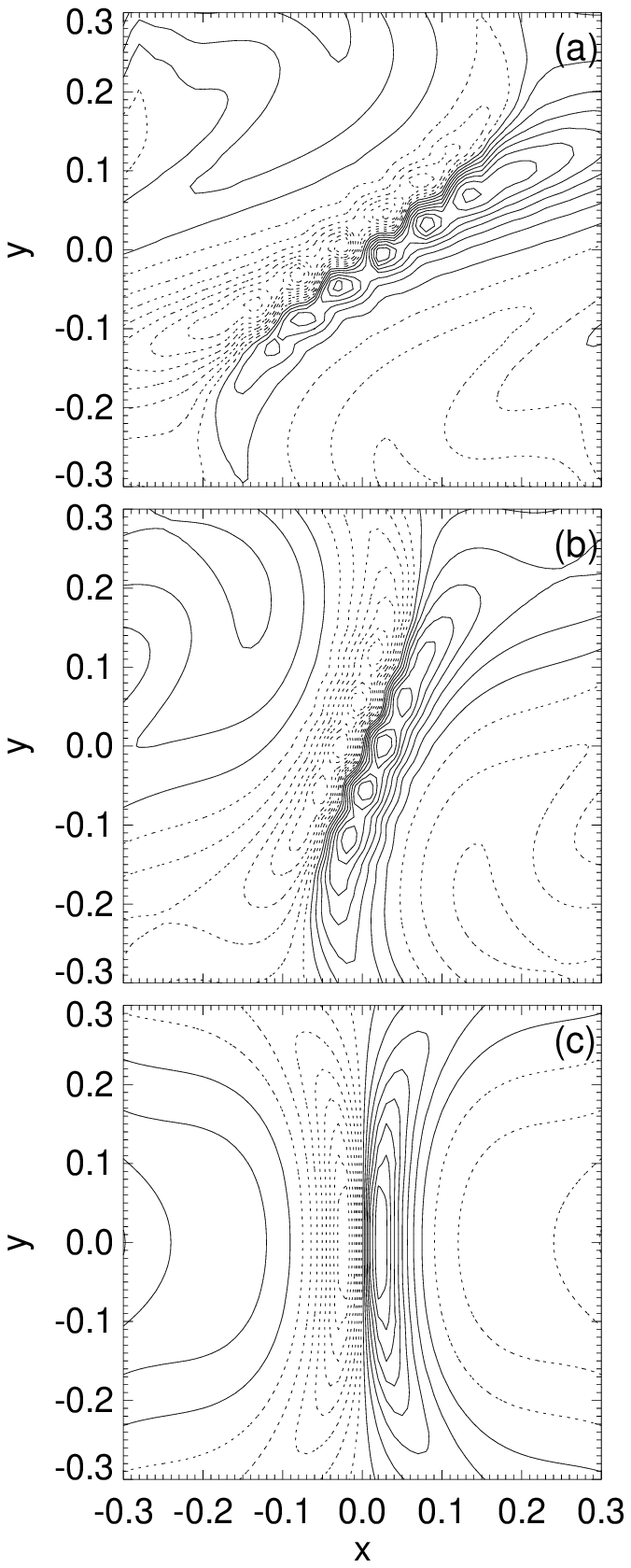]
{ Contours of the current $j_y$ in the planes
 (a) $z=-0.5$, (b) $z=-0.25$, and (c) $z=0$.
 \label{jy} }

\figcaption[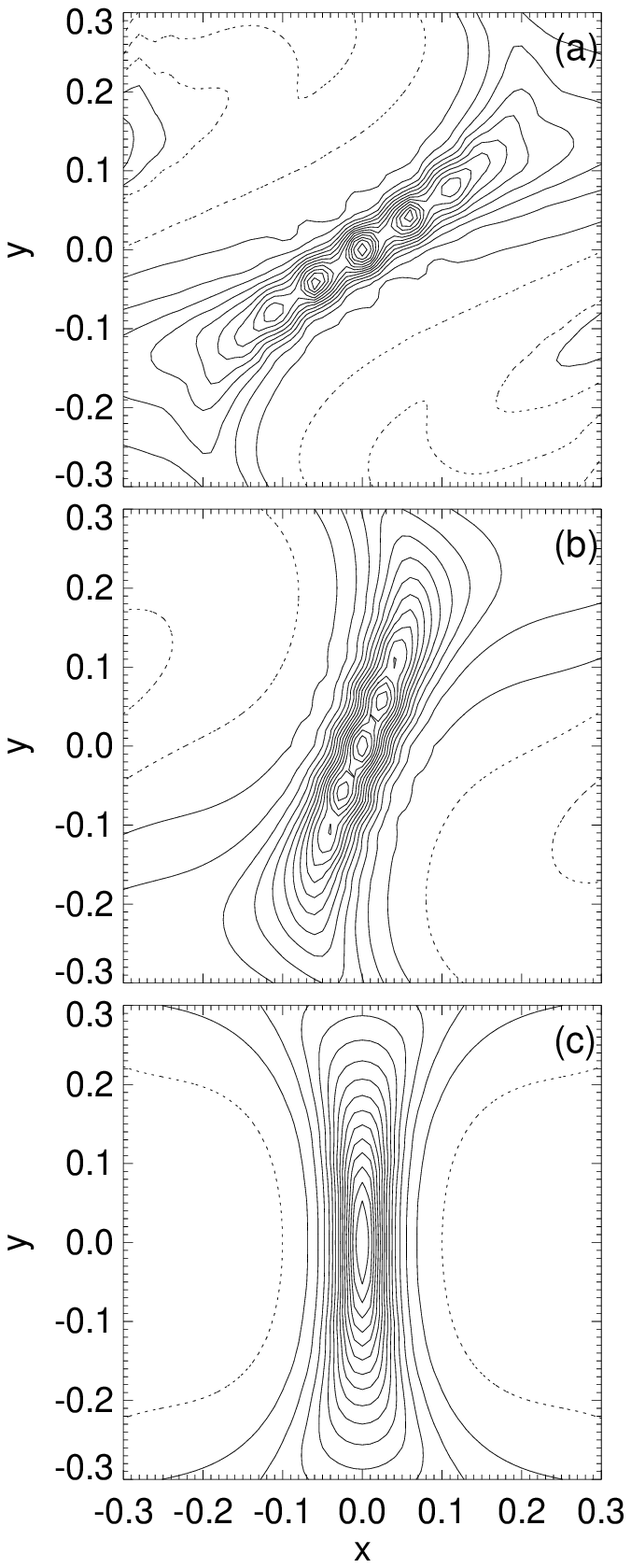]
{ Contours of the current $j_z$ in the planes
 (a) $z=-0.5$, (b) $z=-0.25$, and (c) $z=0$.
 \label{jz} }

\figcaption[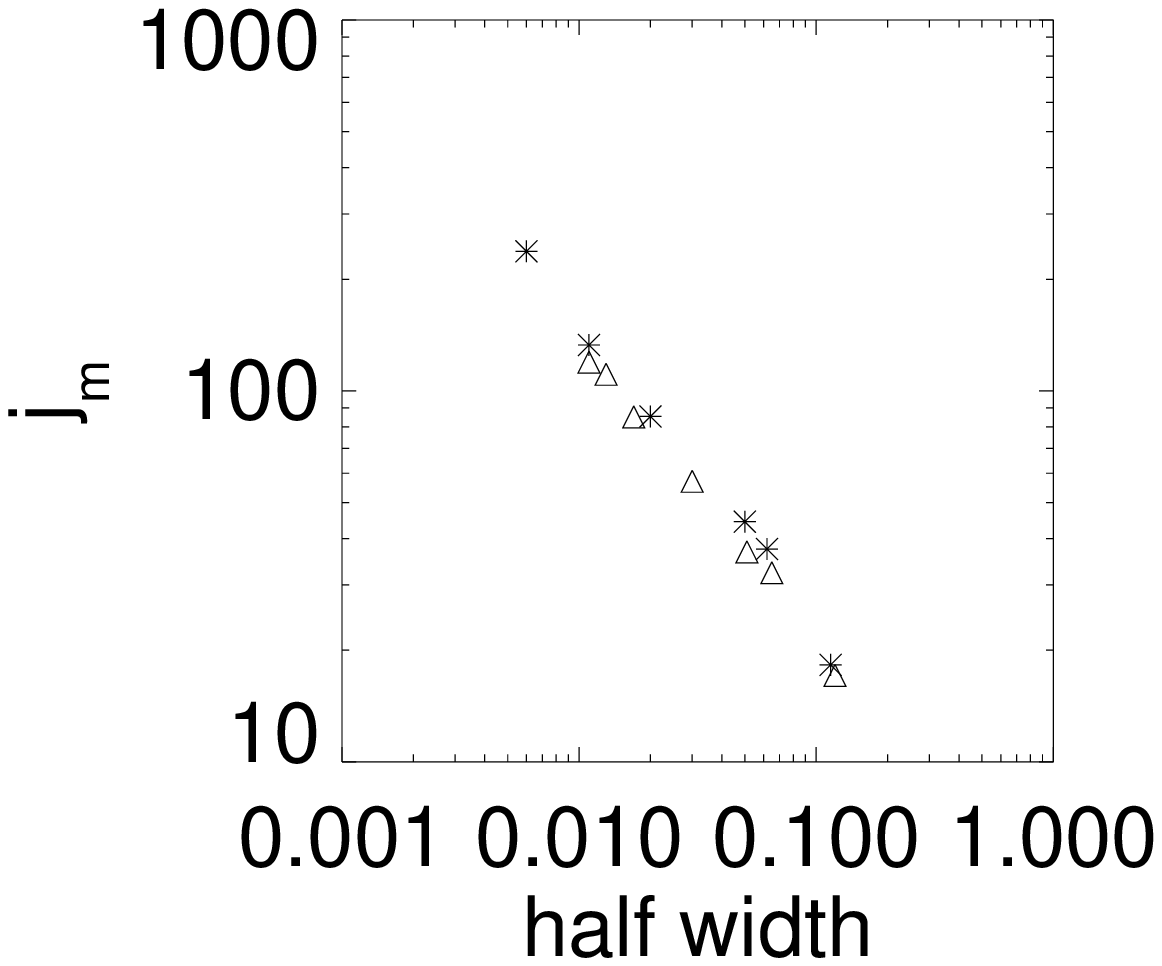]
{ Maximum current against half-width for a varying number of
 grid points. Only the two second shears, 0.6 ($\triangle$)
 and 0.7 ($\times$) that have diverging currents are shown.
 \label{cur_half} }

\figcaption[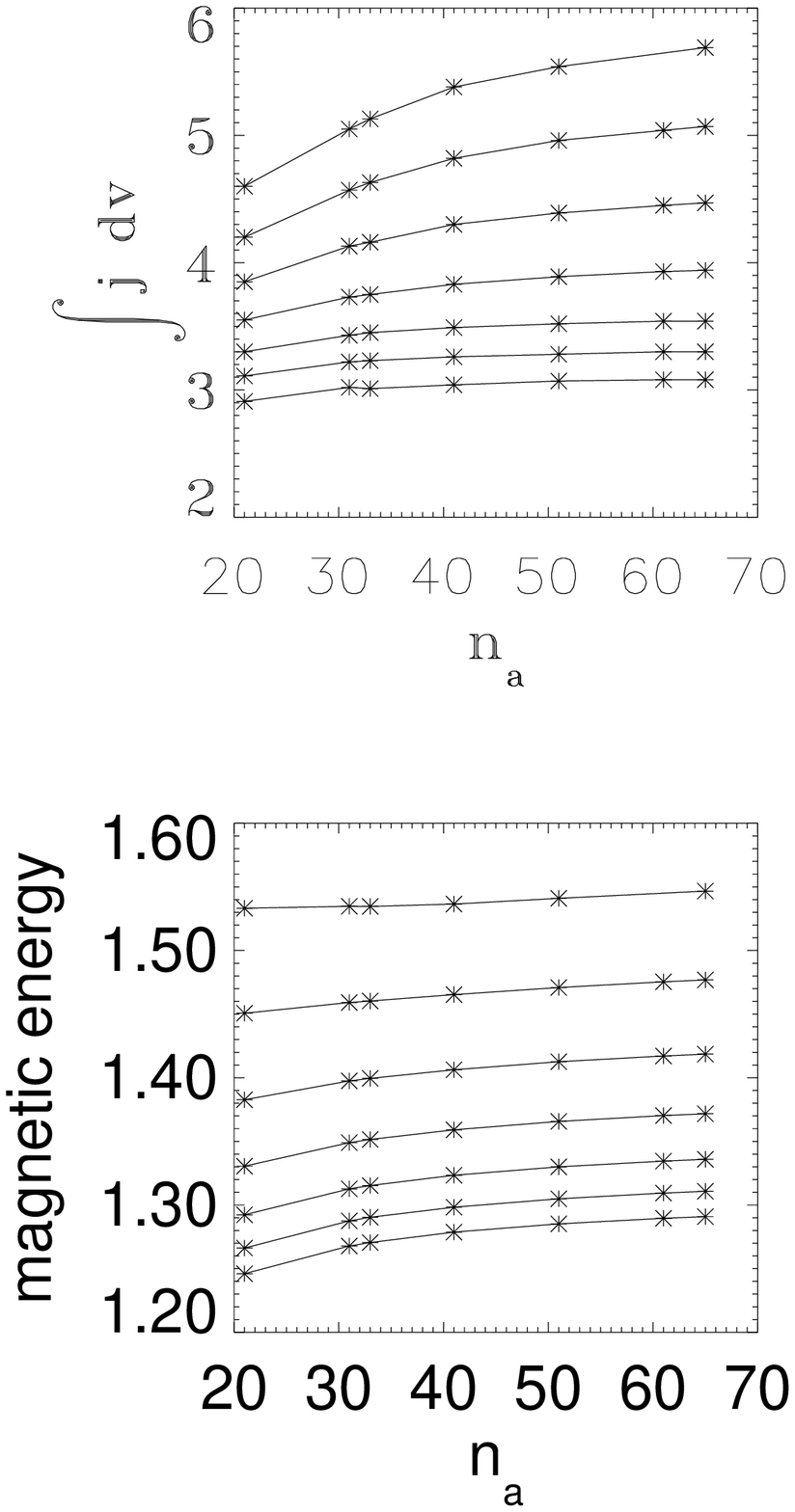]
{ The convergence of the total integrated current and magnetic
 energy as the number of grid points are increased.
 \label{conv} }

\figcaption[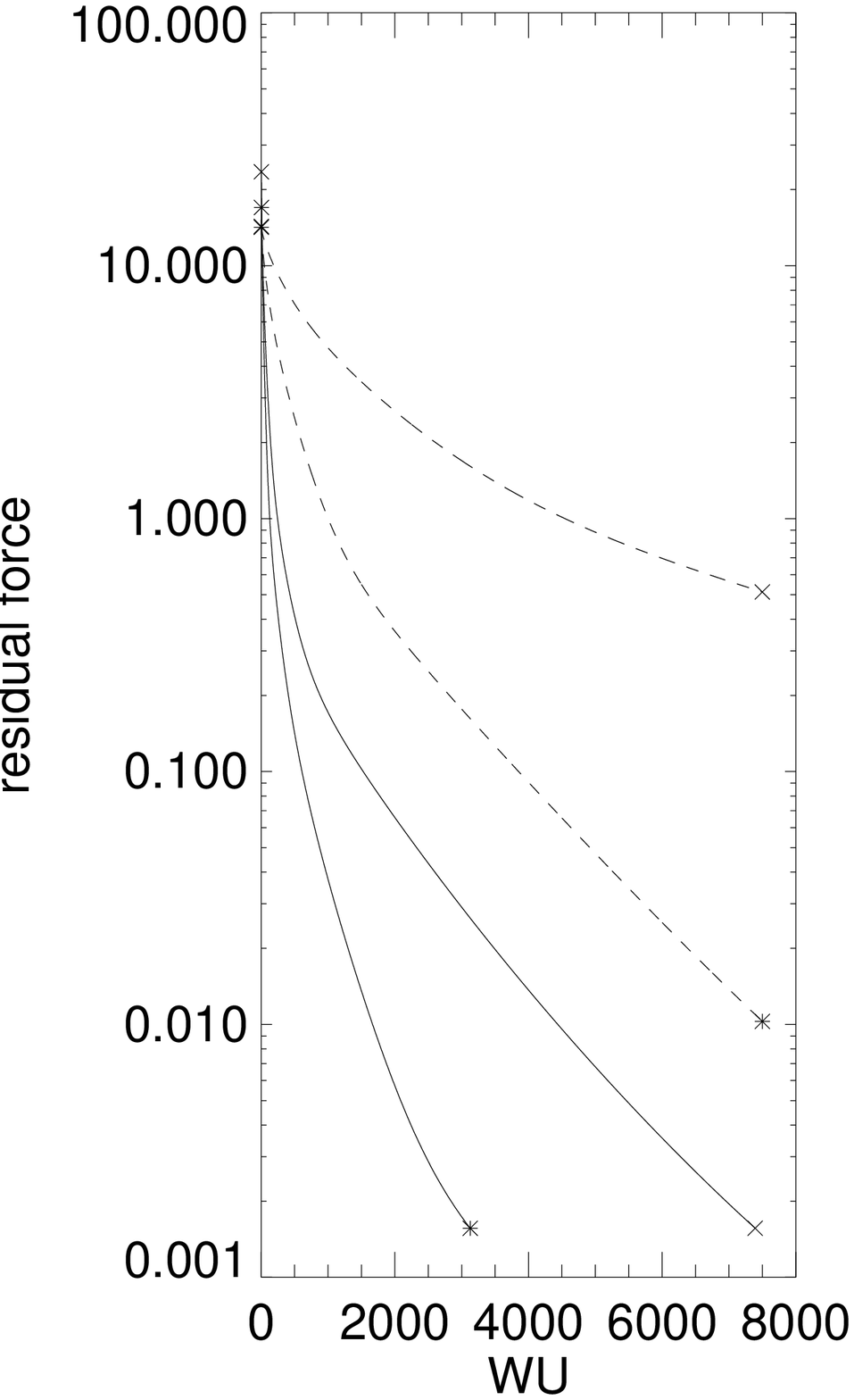]
{ Residual force against WU for a shear of 0.2 with
 $32^3$ ($\ast$) and $64^3$ ($\times$)
 intervals, respectively.
 Dashed curves show the classical method, solid
 curves nonlinear multigrid.
 \label{multigrid} }

\end{document}